\documentclass[a4paper,fleqn,usenatbib]{mnras}

\usepackage{newtxtext,newtxmath}

\usepackage[T1]{fontenc}
\usepackage{ae,aecompl}

\usepackage{graphicx}	
\usepackage{xcolor}           
\usepackage{xfrac}
\usepackage{gensymb}
\usepackage{pdflscape}

\newcommand{\kms}{km$\,$s$^{-1}$}
\newcommand{\Msol}{M$_\odot$}

\title[WISDOM: The SMBH in NGC 7052]{WISDOM project - VII. Molecular gas measurement of the supermassive black hole mass in the elliptical galaxy NGC 7052}

\author[Mark D. Smith et al.]{
Mark D. Smith,$^{1}$\thanks{E-mail: mark.smith@jesus.ox.ac.uk}
Martin Bureau,$^{1,2}$
Timothy A. Davis,$^{3}$
Michele Cappellari,$^{1}$
\newauthor{
Lijie Liu,$^{1}$
Kyoko Onishi,$^{4}$
Satoru Iguchi,$^{5,6}$
Eve V. North,$^{3}$
Marc Sarzi$^{7}$}
\newauthor{
and Thomas G. Williams$^{3,8}$}
\\
$^{1}$Sub-department of Astrophysics, Department of Physics, University of Oxford, Denys Wilkinson Building, Keble Road, Oxford, OX1 3RH, UK\\
$^{2}$Yonsei Frontier Lab and Department of Astronomy, Yonsei University, 50 Yonsei-ro, Seodaemun-gu, Seoul 03722, Republic of Korea\\
$^{3}$School of Physics \& Astronomy, Cardiff University, Queens Buildings, The Parade, Cardiff, CF24 3AA, UK\\
$^{4}$Department of Space, Earth and Environment, Chalmers University of Technology, Onsala Observatory, SE-439 92 Onsala, Sweden\\
$^{5}$Department of Astronomical Science, SOKENDAI (The Graduate University of Advanced Studies), Mitaka, Tokyo 181-8588, Japan\\
$^{6}$National Astronomical Observatory of Japan, National Institutes of Natural Sciences, Mitaka, Tokyo, 181-8588, Japan\\
$^{7}$Armagh Observatory and Planetarium, College Hill, Armagh, BT61 DG, UK\\
$^{8}$Max Planck Institut f\"ur Astronomie, K\"onigstuhl 17, 69117 Heidelberg, Germany\\
}

\date{Accepted 2021 March 15. Received 2021 March 15; in original form 2020 April 14}

\pubyear{2021}

\begin{document}
\label{firstpage}
\pagerange{\pageref{firstpage}--\pageref{lastpage}}
\maketitle

\begin{abstract}
Supermassive black hole (SMBH) masses can be measured by resolving the dynamical influences of the SMBHs on  tracers of the central potentials. Modern long-baseline interferometers have enabled the use of molecular gas as such a tracer. We present here Atacama Large Millimeter/submillimeter Array observations of the elliptical galaxy \mbox{NGC 7052} at $0\farcs11$ ($37\,$pc) resolution in the $^{12}$CO(2-1) line and $1.3\,$mm continuum emission. This resolution is sufficient to resolve the region in which the potential is dominated by the SMBH. We forward model these observations, using a multi-Gaussian expansion of a \textit{Hubble Space Telescope} F814W image and a spatially-constant mass-to-light ratio to model the stellar mass distribution. We infer a SMBH mass of $2.5\pm0.3\times10^{9}\,\mathrm{M_\odot}$ and a stellar \textit{I}-band mass-to-light ratio of $4.6\pm0.2\,\mathrm{M_\odot/L_{\odot,I}}$ ($3\sigma$ confidence intervals). This SMBH mass is significantly larger than that derived using ionised gas kinematics, which however appear significantly more kinematically disturbed than the molecular gas. We also show that a central molecular gas deficit is likely to be the result of tidal disruption of molecular gas clouds due to the strong gradient in the central gravitational potential.
\end{abstract}




\section{Introduction}
Supermassive black holes (SMBHs) are characterised by just a few properties: their masses, spins and charges. A SMBH mass can be measured by spatially- and/or temporally-resolving a dynamical tracer of the central potential, most commonly stars or ionised gas (see \citealt{Kormendy+1995ARAA33.581} for a review contrasting these two methods), or more rarely masers (e.g. \citealt{Miyoshi+1995Nature373.127, Greenhill+1995AandA304.21}). The last three decades of studies have demonstrated that SMBH masses correlate tightly with a wide variety of properties of their host galaxies, including the stellar velocity dispersion \citep[e.g.][]{Gebhardt+2000ApJL539.13, Ferrarese+2000ApJL539.9}, bulge mass and/or luminosity \citep[e.g.][]{Kormendy+1995ARAA33.581, Magorrian+1998AJ115.2285}, total luminosity \citep[e.g.][]{Kormendy+2001AIPC586.363} and S\'ersic index \citep[e.g.][]{Graham+2001ApJL563.11}. These correlations are sufficiently tight to imply (potentially self-regulating) co-evolutionary processes. That the tightest correlations are found with the properties of classical (merger-formed) bulges \citep[e.g.][]{Gultekin+2009ApJ698.198, Beifiori+2012MNRAS419.2497, vdBosch2016ApJ831.134, Saglia+2016ApJ818.47} suggests that mergers may be important (either via the SMBHs themselves merging or by the disrupted potential leading to enhanced accretion onto a central SMBH; e.g. \citealt{Sanders+1988ApJ325.74, Hernquist1989Nature340.687, DiMatteo+2005Nature433.604}). However, the potential importance of secular accretion onto a SMBH, coupled with galactic evolution at larger spatial scales via active galactic nucleus (AGN) feedback, cannot be discounted. Simulations indicate that such feedback is vital for replicating observed properties on large scales \citep[e.g.][]{Benson+2003ApJ599.38,McNamara+2007ARAA45.117}. Nevertheless, the relative importance of these processes remains disputed \citep[e.g.][]{Kormendy+2013ARAA51.511, Simmons+2017MNRAS470.1559, Krajnovic+2018MNRAS473.5237}.

Molecular gas emission has proved to be a suitable tracer of SMBH potentials \citep[e.g.][]{Davis+2013Nature494.328} for galaxies across the Hubble sequence, including those hosting AGN. Our millimetre-Wave Interferometric Survey of Dark Object Masses (WISDOM) exploits the high angular resolution available from modern interferometers to spatially-resolve CO emission on SMBH-dominated scales. In previous papers in this sequence, we have presented new SMBH measurements \citep{Davis+2017MNRAS468.4675, Davis+2018MNRAS473.3818, Onishi+2017MNRAS468.4663, Smith+2019MNRAS485.4359, North+2019MNRAS490.319}, explored a correlation between CO line width and SMBH mass \citep{Smith+2020MNRASsubmitted}, and studied the properties of the cold molecular interstellar medium at very high resolution in local galaxies \citep{Liu+2020MNRASinrev}. In parallel, other groups have used this technique to measure SMBH masses in various other galaxies \citep[e.g.][]{Barth+2016ApJL822.28, Barth+2016ApJ823.51, Boizelle+2019ApJ881.10, Nagai+2019arXiv190506017, Thater+2020IAUS353.199}. Notably, using this method, robust constraints have even been placed on a few SMBH masses in dwarf galaxies \citep{Nguyen+2020ApJ892.68, Davis+2020MNRAS496.4061}.

In this paper, we use new high-resolution observations of the galaxy \mbox{NGC 7052} to measure its central SMBH mass. In Section \ref{sec_NGC7052}, we describe the properties of our target galaxy. Section \ref{sec_observations} describes the Atacama Large Millimeter/submillimeter Array (ALMA) observations, their calibration and imaging. The dynamical model we fit to our observations is described in Section \ref{sec_dynamics}, and we discuss our results in Section \ref{sec_discussion}. We conclude briefly in Section \ref{sec_conclusions}. Throughout this paper, velocities are given in the radio convention.

\section{NGC 7052}
\label{sec_NGC7052}
\begin{figure*}
	\centering
	\includegraphics[trim={2.5cm 2cm 2.5cm 2.5cm},width=0.7\textwidth]{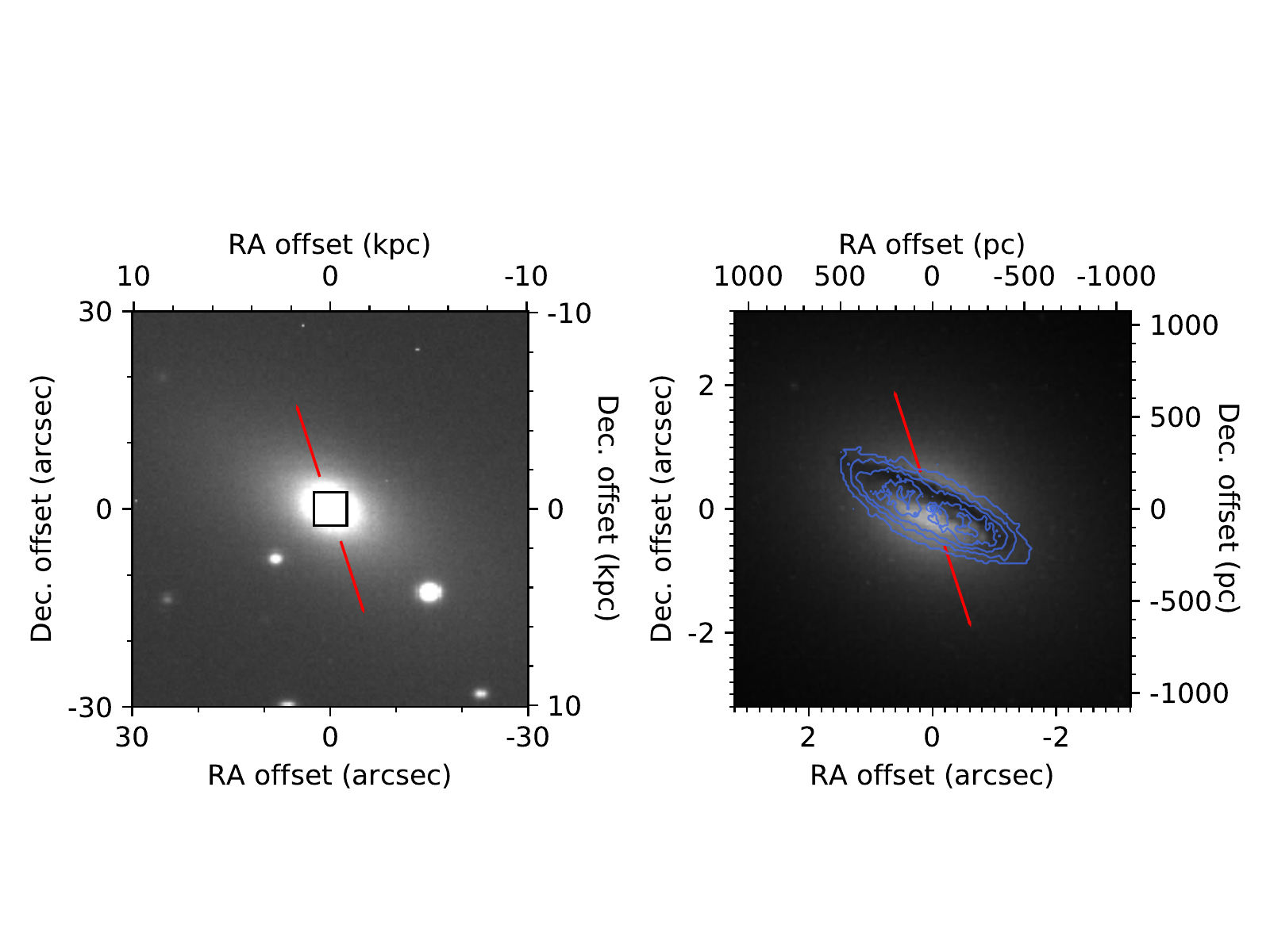}
         \vspace{0.3cm}
         \caption{\textbf{Left panel:} Jakobus Kapteyn Telescope \textit{V}-band image of \mbox{NGC 7052} (greyscale), showing the large-scale morphology of \mbox{NGC 7052}. The black central box is the area shown in the right panel. \textbf{Right panel:} Unsharp-masked \textit{HST} WFPC2/PC F814W image of \mbox{NGC 7052} (greyscale; \protect\citealt{vanderMarel+1998AJ116.2220}), showing the central dust disc. Overlaid are the H$_2$ surface density contours (blue) inferred from our ALMA observations, assuming a CO(2-1)/CO(1-0) line ratio of unity and $\alpha_\mathrm{CO}=4.3\,$\Msol$\,$(K\,\kms)$^{-1}\,$pc$^{-2}$. The contours are from the level at which the noise was clipped, $15\,$\Msol$\,$pc$^{-2}$, and then at $4000$, $8000$, $12\,000$, and $16\,000\,$\Msol$\,$pc$^{-2}$. Spatial offsets are relative to the $1.3\,$mm continuum source position listed in Table \ref{tab_continuum}. The red lines in each panel indicate the $6\,$cm radio emission axis \protect\citep{Condon+1991AJ101.362}.}
    \label{fig_NGC7052_optical}
\end{figure*}

\mbox{NGC 7052} is an isolated elliptical radio galaxy (Figure \ref{fig_NGC7052_optical}, left panel) in the Vulpecula constellation, located at \mbox{$21^\rmn{h}18^\rmn{m}33^\rmn{s}$},\,\mbox{$+26\degree26\arcmin49\arcsec$}. Its total stellar mass is $5.6\times10^{11}\,$\Msol$\,$\citep{Pandya+2017ApJ837.40}, among the most massive galaxies in the local universe, and it is a member of the MASSIVE sample of such galaxies \citep{Ma+2014ApJ795.158}. The near-infrared effective (i.e.$\,$half-light) radius ($R_\mathrm{e}$) is $14\farcs7$ \citep{Ma+2014ApJ795.158}. The galaxy is kinematically classified as a slow-rotator according to the criterion of \cite{Emsellem+2011MNRAS414.888}, based on the projected stellar angular momentum $\lambda_\mathrm{e}$ (spin parameter) averaged within one effective radius ($\lambda_\mathrm{e}=0.15$; \citealt{Veale+2017MNRAS471.1428}). Throughout this paper, we adopt the distance used in the MASSIVE survey, $D=69.3\,$Mpc. This distance is calculated from the observed recession velocity and the flow model of \cite{Mould+2000ApJ529.786} assuming a current Hubble constant $H_0=70\,$\kms$\,$Mpc$^{-1}$. At this distance, $1\arcsec$ corresponds to $336\,$pc.

Radio jets have been mapped in \mbox{NGC 7052} on arc-minute scales at $1.5$ and $5\,$GHz using the Very Large Array (VLA; \citealt{Parma+1986AandAS64.135}) and Westerbork Synthesis Radio Telescope (WSRT; \citealt{Fanti+1977AandAS29.279}), respectively. The radially-declining profile of this emission indicates the galaxy is a Fanaroff-Riley Class I source \citep[FR-I;][]{Capetti+2000MNRAS318.493, Capetti+2002AandA383.104}. 

X-ray emission from the galaxy has been detected and extensively studied \citep[e.g.][]{Donato+2004ApJ617.915, Mulchaey+2010ApJL715.1, Goulding+2016ApJ826.167}. \cite{Memola+2009AandA497.359} used \textit{Chandra} observations to separate the contribution of the AGN from that of the spatially-unresolved X-ray binaries, determining an AGN X-ray luminosity of $L_\mathrm{AGN, X}{\approx}10^{33}\,$W. 

Optical images from the \textit{Hubble Space Telescope} (\textit{HST}) reveal that the centre of \mbox{NGC 7052} harbours a prominent nuclear dust disc with a dust mass of ${\approx}10^4\,\mathrm{M_\odot}$ \citep{Nieto+1990AandA235.L17}, shown here in extinction in Figure \ref{fig_NGC7052_optical} (right panel). This dust disc has a semi-major (-minor) axis of $1\farcs94$ ($0\farcs67$); assuming the dust disc has no intrinsic thickness yields an inclination estimate of $70\pm2\degree$ \citep{vanderMarel+1998AJ116.2220}. Although the dust disc is very prominent to the north-west of the nucleus, it does not appear to significantly obscure the nucleus itself \citep{Capetti+2000MNRAS318.493}. It is not orthogonal to the radio emission \citep{Capetti+1999MNRAS304.434}. 

Despite being an early-type galaxy, \mbox{NGC 7052} hosts a significant molecular gas reservoir with a total mass of $2.3\times10^9\,\mathrm{M_\odot}\,$\citep[][corrected to $\alpha_\mathrm{CO}=4.3\,$\Msol$\,$(K\,\kms)$^{-1}\,$pc$^{-2}$]{Wang+1992AJ104.2097}. Warm gas makes up only a very small proportion of the galaxy's mass budget, totalling only $4\times10^3\,$\Msol$\,$(estimated from the H$\beta$ luminosity) over the central $1.7\,$kpc radius \citep{Pandya+2017ApJ837.40}. 

\textit{HST} Faint Object Spectrograph (FOS) observations of the H$\alpha$ and [\ion{N}{II}] emission lines along the major axis were modelled by \cite{vanderMarel+1998AJ116.2220} to determine a central SMBH mass of $M_\mathrm{BH}=3.9^{+2.7}_{-1.5}\times10^8\,$\Msol$\,$(corrected to our adopted distance), robustly excluding models without a central SMBH. However, the ionised gas kinematics in the centre of the galaxy are dominated by turbulent motions (exceeding $400\,$\kms), the potential dynamical support of which were neglected by \cite{vanderMarel+1998AJ116.2220}. This likely leads to an underestimated SMBH mass, as seen for example in Centaurus A \citep{HaringNeumayer+2006ApJ643.226}. Cold molecular gas generally exhibits smaller velocity dispersions, allowing an independent, and likely more reliable, dynamical SMBH mass measurement. While the most precise SMBH mass measurements so far have been achieved by tracing maser emission very close to the SMBHs with very long baseline interferometry (VLBI; e.g. \citealt{Miyoshi+1995Nature373.127, Kuo+2011ApJ727.20, Gao+2017ApJ834.52}), no $22\,$GHz maser emission was detected in NGC~7052 with the Effelsberg 100-m telescope \citep{Braatz+1996ApJS106.51}, leaving cold molecular gas the most promising option.

The sphere of influence of the SMBH, the approximate physical scale at which the SMBH dominates the gravitational potential, is given by $R_\mathrm{SoI}\equiv GM_\mathrm{BH}/\sigma_\mathrm{e}^2$, where $G$ is the gravitational constant, $M_\mathrm{BH}$ the SMBH mass and $\sigma_\mathrm{e}$ the stellar velocity dispersion averaged within $1\,R_\mathrm{e}$. Using the distance-corrected SMBH mass from \cite{vanderMarel+1998AJ116.2220} and $\sigma_\mathrm{e}=266\pm13\,$\kms \citep{Gultekin+2009ApJ698.198}, we estimate $R_\mathrm{SoI} = 24\pm13\,$pc ($0\farcs07\pm0\farcs04$).

\section{ALMA observations}
\label{sec_observations}
\mbox{NGC 7052} was observed with the ALMA 12-m array as part of the WISDOM project 2018.1.00397.S. An extended ALMA configuration was used to provide baselines of $40\,$m--$5.9\,$km, in two tracks on 8th and 9th August 2018, each on-source for 21 minutes. The former track failed the on-line ALMA quality assessment check (known as QA0) due to large residuals in the phase calibration, and therefore the second track was taken. Manual calibration was performed on the first track by the United Kingdom ALMA Regional Centre, recovering much of the data for further use. The second track was automatically calibrated by the ALMA pipeline, and one antenna (DA45) was subsequently manually flagged due to an amplitude error.

To better sample the \textit{uv} plane and thus recover any large-scale structure, additional observations were taken with a compact ALMA configuration and with the 7-m Atacama Compact Array (ACA). The additional 12-m track on 31st October 2018 covered baselines $15\,$m--$1.4\,$km and was on-source for 5 minutes. The ACA track was obtained as part of programme 2016.2.00046.S, was observed on 21st August 2019, covered baselines $9\,$--$\,45\,$m, and was on-source for 32 minutes. Both of these tracks were automatically calibrated by the ALMA pipeline.

The properties of these four observing tracks are listed in Table~\ref{tab_observingTracks}. Combining all four tracks together yields continuous baseline coverage from $9\,$m to $5.9\,$km, corresponding to sensitivity to angular scales from $0\farcs06$ to $36\arcsec$. The dust disc visible in optical images of \mbox{NGC 7052} has a major- (minor-)axis diameter of ${\approx}4\arcsec$ (${\approx}1\arcsec$). Assuming the CO is co-spatial with the dust disc, we therefore expect to recover all the emitted flux. 

\begin{table*}
	\centering
	\caption{Properties of the four observing tracks.}
	\label{tab_observingTracks}
	\begin{tabular}{lccccl} 
		\hline
		Track & Date & Array & Baseline range & On-source time  & Calibration\\
		\hline
		\texttt{uid\_A002\_Xc39302\_X5d57} & 21st \phantom{0\!}August 2017 & \phantom{0}7-m & $9\,$m--$45\,$m & $32\,$min & Pipeline\\
		\texttt{uid\_A002\_Xd44a99\_X974} & 31st October 2018 & 12-m & $15\,$m--$1.4\,$km & $\phantom{0}5\,$min & Pipeline\\
		\texttt{uid\_A002\_Xdfcc3f\_X1c7a} & \phantom{0}8th \phantom{0\!}August 2019 & 12-m & $40\,$m--$5.9\,$km & $21\,$min & Manual\\
		\texttt{uid\_A002\_Xdfdbea\_X598} & \phantom{0}9th \phantom{0\!}August 2019 & 12-m & $40\,$m--$5.9\,$km & $21\,$min & Pipeline; antenna DA45 flagged\\
		\hline
	\end{tabular}
\end{table*}

Two spectral setups were used. For all 12-m array observations, a $1.875\,$GHz bandwidth spectral window with a channel width of ${\approx}1\,$MHz was placed over the $^{12}$CO(2-1) emission line. At this frequency, this corresponds to a ${\approx}2400\,$km$\,$s$^{-1}$ velocity range and ${\approx}1\,$km$\,$s$^{-1}$ channels. The ACA observations used a slightly different receiver configuration, with one $2\,$GHz (${\approx}2600\,$km$\,$s$^{-1}$) bandwidth spectral window and $500\,$kHz (${\approx}0.7\,$km$\,$s$^{-1}$) channels. In both cases, the remaining three $2\,$GHz bandwidth spectral windows were placed to detect continuum emission.

\subsection{Continuum images}
The calibrated observations were concatenated (with default weighting and the \texttt{concat} task) using the \texttt{Common Astronomy Software Applications (CASA)} package \citep{McMullin+2007ASPC376.127}, and an image of the $1.3\,$mm continuum was created using the \texttt{CASA} task \texttt{tclean} in multi-frequency synthesis mode. The continuum spectral windows and line-free channels of the line spectral window were used. The image was made using Briggs weighting with a robust parameter of 0, balancing angular resolution and sensitivity. An approximately point-like continuum source was detected and fit with a two-dimensional (2D) Gaussian using the \texttt{CASA} task \texttt{imfit}. The properties of this continuum image and of the detected continuum source are listed in Table \ref{tab_continuum}.

\begin{table}
\centering
	\caption{Parameters of the continuum image and the detected $1.3\,$mm continuum source.}
	\label{tab_continuum}
	\begin{tabular}{lc} 
		\hline
		Image property & Value\\
		\hline
		Image size (pix) & $512\times512$\\
		Image size (arcsec) & $10.24\times10.24$\\
		Image size (pc) & $3440\times3440$\\
		Pixel scale (arcsec$\,$pix$^{-1}$) & 0.02\\
		Pixel scale (pc$\,$pix$^{-1}$) & 6.72\\
		$1\sigma\,$ sensitivity ($\upmu$Jy$\,$beam$^{-1}$) & 80\\
		Synthesised beam (arcsec) & $0.12\times0.09$\\
		Synthesised beam (pc) & $38\times27$\\
		\hline
		Source property & Value \\
		\hline
		Right ascension & $21^\rmn{h}18^\rmn{m}33\fs0433\pm0\fs0001$\\
		Declination & ${+}26\degree26\arcmin49\farcs242\pm0\farcs003$\\
		Integrated flux (mJy) & $22.3\pm1.5$\\
		Deconvolved size (arcsec) & $(0.07\pm0.02)\times(0.05\pm0.03)$\\
		Deconvolved size (pc) & $(22\pm6)\times(16\pm9)$\\
		\hline
	\end{tabular}
\end{table}

\subsection{Line images}
A linear fit to the continuum spectral windows and line-free channels of the line spectral window was subtracted from the \textit{uv}-plane data using the \texttt{CASA} task \texttt{uvcontsub}. The continuum-subtracted data were then concatenated, imaged and cleaned using the `cube' mode of the \texttt{tclean} task and adopting Briggs weighting with robust=0. The properties of the resulting image cube are listed in Table \ref{tab_COdata}.

\begin{table}
\centering
	\caption{Parameters of the CO line cube.}
	\label{tab_COdata}
	\begin{tabular}{lc} 
		\hline
		Image property & Value\\
		\hline
		Image size (pix) & $512\times512$\\
		Image size (arcsec) & $10.24\times10.24$\\
		Image size (pc) & $3440\times3440$\\
		Pixel scale (arcsec$\,$pix$^{-1}$) & $0.02$\\
		Pixel scale (pc$\,$pix$^{-1}$) & 6.72\\
		Velocity range (\kms) & $4035-5235$\\
		Channel width (\kms) & $15$\\
		$1\sigma\,$ sensitivity (mJy$\,$beam$^{-1}$) & 0.5\\
		$1\sigma\,$ sensitivity (\Msol$\,$pc$^{-2}$) & 15\\
		Synthesised beam (arcsec) & $0.13\times0.10$\\
		Synthesised beam (pc) & $41\times30$\\
		\hline
	\end{tabular}
\end{table}

\begin{landscape}
\begin{figure} \begin{center}
\centering
	\includegraphics[scale=1.16,trim={0cm 0.4cm 0.6cm 0.6cm}]{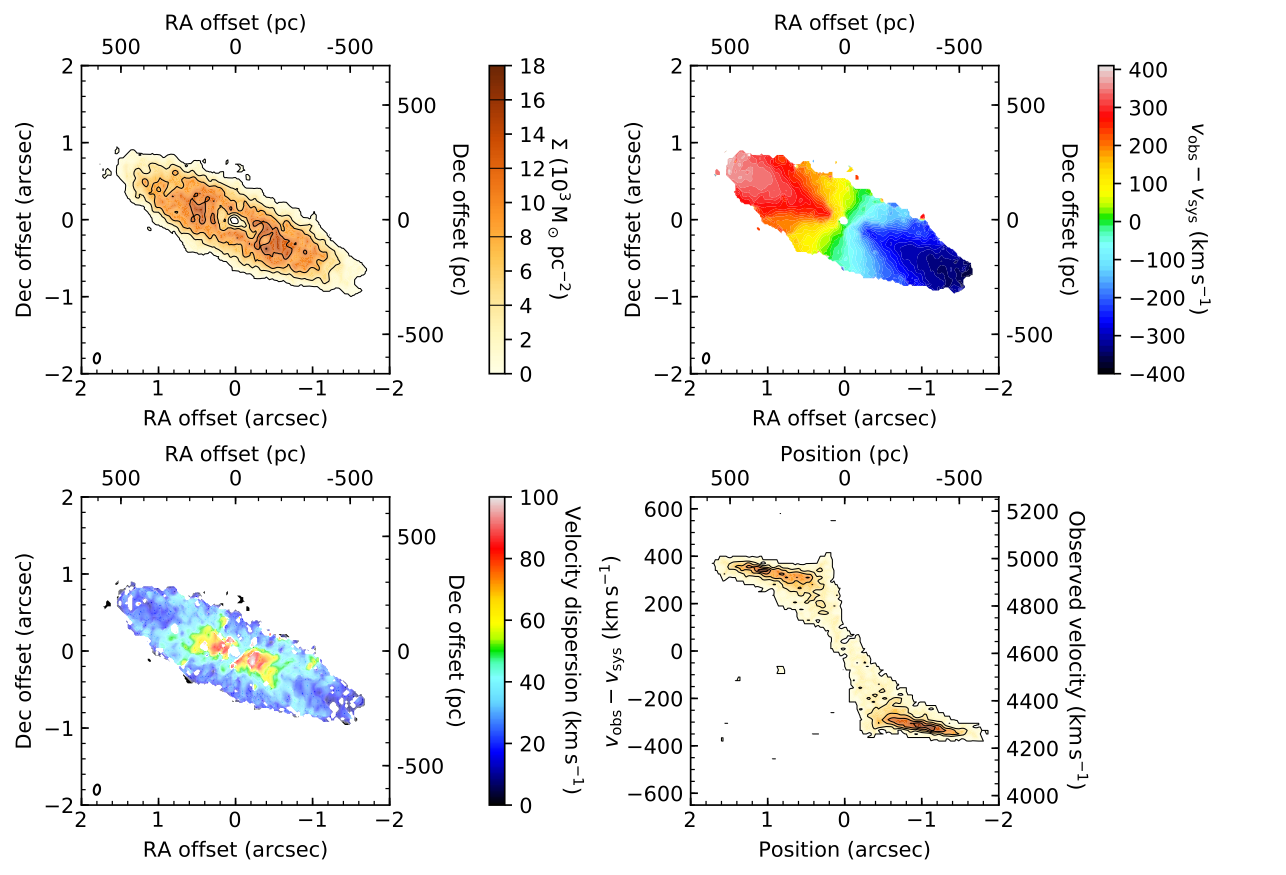}
         
         \caption{Moment maps of the $^{12}$CO(2-1) emission in \mbox{NGC 7052} centred on the compact continuum source. \textbf{Top-left:} Molecular gas surface density (orange scale and black contours), assuming a CO(2-1)/CO(1-0) line ratio of unity and $\alpha_\mathrm{CO}=4.3\,$\Msol$\,$(K\,\kms)$^{-1}\,$pc$^{-2}$. Black contours are from the level at which the noise was clipped, $15\,$\Msol$\,$pc$^{-2}$, and then at $4000$, $8000$, $12\,000$ and $16\,000\,$\Msol$\,$pc$^{-2}$. \textbf{Top-right:} Mean line-of-sight velocity. \textbf{Bottom-left:} Line-of-sight velocity dispersion. \textbf{Bottom-right:} Kinematic major-axis position-velocity diagram (PVD; orange scale and black contours). In both right panels, $v_\mathrm{obs}$ is the observed line-of-sight velocity and $v_\mathrm{sys}=4610\,$\kms$\,$is the galaxy systemic velocity in the radio convention. The maps show the synthesised beam in their bottom-left corners.}
    \label{fig_NGC7052_maps}
\end{center}
\end{figure}
\end{landscape}

The molecular gas distribution, mean line-of-sight velocity field, velocity dispersion field, and kinematic major-axis position-velocity diagram (PVD) are shown in Figure \ref{fig_NGC7052_maps}. These were made with the masked-moments method \citep{Dame2011arXiv1101.1499}, whereby the cube is first convolved spatially by the beam and Hanning-smoothed spectrally, pixels that exceed a noise threshold are included in a mask, and this mask is then applied to the original cube. This method selects only areas of structured emission in the original cube and excludes regions with no significant emission, thus producing improved moment maps. The spectrally-integrated intensity map is then converted into molecular gas surface densities by appropriately modifying Equation 3 of \cite{Bolatto+2013ARAA51.207}, and adopting a CO(2-1)/CO(1-0) line ratio of unity and $\alpha_\mathrm{CO}=4.3\,$\Msol$\,$(K\,\kms)$^{-1}\,$pc$^{-2}$.

The CO gas in \mbox{NGC 7052} is distributed in a regularly rotating disc, coincident with the dust disc (Figure \ref{fig_NGC7052_optical}, right panel). The total molecular gas mass derived from our data is $1.8\times10^9\,\mathrm{M_\odot}$. This is very similar to the single-dish measurement of \cite{Wang+1992AJ104.2097}, and likely fully consistent once the systematic uncertainties on the absolute flux calibrations of ALMA and the Nobeyama 45-m telescope are considered, further evidence that we have not resolved out significant flux. The CO surface density peaks along the major axis at ${\approx}0\farcs5$ on either side of the centre, rapidly decreasing toward the nucleus and more slowly outward. In the very centre of the galaxy is a small hole, where the gas surface density is below our sensitivity limit of $15\,$\Msol$\,$pc$^{-2}$. 

In principle, such a hole could be an artefact caused by projecting the cube onto an image. Indeed, the mask could exclude gas close to the SMBH where the line-of-sight velocity distribution becomes very broad, causing emission to be spread over many channels (and therefore fall below our sensitivity limit in any given channel). However, we have checked that a manually-defined mask including all channels within the hole does not recover any more emission. Another possibility is that a few channels showing absorption against the continuum source contribute negative flux in this region (once continuum subtracted), reducing the sum. There is however no evidence of such absorption features in the spectra within the hole. To further exclude the possibility that erroneous continuum-subtraction has created the hole, we made a second data cube from the observations without first subtracting the continuum. The hole was still visible in this cube, the continuum source not being sufficiently extended to fill the void. Having excluded these two explanations, we conclude that the hole is genuine and astrophysical in origin. We discuss it further in Section \ref{ssec_shear}. 

Such holes appear to be common in the galaxies studied in the WISDOM survey. Typically, they have spatial extents similar to those of the SMBH spheres-of-influence, occasionally preventing the detection of the central Keplerian rotation \citep[e.g.][]{Davis+2018MNRAS473.3818,Smith+2019MNRAS485.4359}. In such cases we have nevertheless been able to measure the SMBH masses, as the SMBH's presence still enhances the gas velocities above those expected from the stars alone. 

The kinematic major-axis PVD (Figure \ref{fig_NGC7052_maps}, bottom-right panel) shows a rotation curve that rises towards the centre with decreasing radius at radii $r<0\farcs5$, as would be expected from Keplerian rotation around a compact mass. The signature is most prominent on the north-east side of the galaxy (positive velocities), albeit only in the faintest contour, while it is only marginally visible on the south-west side (negative velocities), due to the slight asymmetry of the CO disc. Additional evidence for the enhanced velocities due to the presence of a central mass concentration is given by the shape of the PVD envelope. The gas remains at high velocities to very small radii (${\approx}250\,$\kms$\,$ at $0\farcs2$ or $70\,$pc), before falling very steeply. In the absence of a central mass concentration, a shallower central decline would be expected.

The velocity dispersion map (Figure \ref{fig_NGC7052_maps}, bottom-left panel) indicates that the gas at the edge of the disc is dynamically cold ($\sigma_\mathrm{gas}{<}30\,$\kms). As the gas density increases, the dispersion also increases, but in the centre of the disc it is likely that the line-of-sight velocity dispersion is dominated by (beam) smearing of closely-spaced isovelocity contours. This suggests that the molecular gas remains dynamically cold throughout the disc, in contrast to the strong gradients observed in ionised gas \citep{vdBosch+1995MNRAS274.884}. We will further test this conclusion using our dynamical modelling in Section \ref{sec_dynamics}.

\section{Dynamical modelling}
\label{sec_dynamics}
Dynamical modelling of NGC 7052 was carried out using the same methods as extensively discussed in previous works of this series (particularly \citealt{Davis+2017MNRAS468.4675} and \citealt{Smith+2019MNRAS485.4359}), so we provide only an outline of our procedures here, before discussing in greater detail features of the model unique to this case.

Simulated data cubes were constructed from dynamical models of the molecular gas disc in NGC 7052 using the \texttt{Integrated Development Language (IDL)} version of the \texttt{Kinematic Molecular Simulation (KinMS)} tool\footnote{\href{https://github.com/TimothyADavis/KinMS}{https://github.com/TimothyADavis/KinMS}} \citep{Davis+2013MNRAS429.534}. These were fit to the observed data cube using a Markov-chain Monte-Carlo (MCMC) method with a custom Gibbs sampler (\texttt{KinMS\_mcmc}\footnote{\href{https://github.com/TimothyADavis/KinMS\_mcmc}{https://github.com/TimothyADavis/KinMS\_mcmc}}). \texttt{KinMS} generates a set of particles at positions replicating a specified surface brightness profile, it assigns to each particle the velocity expected at its radius from a specified circular velocity curve (although every particle is also assigned an additional random velocity, depending on the velocity dispersion selected by the user, that is not taken into account dynamically), it projects these velocities along the line of sight (according to the specified galaxy viewing angles), and it places the particle into a data cube. This cube is then convolved spatially by the synthesised beam to replicate instrumental effects.

The circular velocity at every radius is calculated (using the \texttt{IDL} procedure \texttt{MGE\_CIRCULAR\_VELOCITY}\footnote{\href{http://purl.org/cappellari/software}{http://purl.org/cappellari/software}}) from the SMBH mass and a model of the stellar mass distribution, parametrized by a multi-Gaussian expansion (MGE; \citealt{Emsellem+1994A&A285.723,Cappellari2002MNRAS333.400}) of a \textit{HST} image and a stellar mass-to-light ratio $M/L$. This stellar contribution is explained in further detail in Section \ref{sec_mge}, listed in Table \ref{tab_mge}, and shown in Figure \ref{fig_mge}. 

In addition to these three dynamical parameters (SMBH mass, stellar mass-to-light ratio and gas velocity dispersion), and two parameters specifying the disc orientation relative to the observer (inclination and position angle), we also allow the model to vary four `nuisance' parameters. The kinematic centre of the galaxy can have small spatial and velocity offsets with respect to the location of the aforementioned continuum source and the galaxy systemic velocity, and we let the surface brightness function have an arbitrary overall scaling.

\subsection{Stellar potential}
\label{sec_mge}
The stellar potential is determined from a \textit{HST} Wide Field Planetary Camera 2 (WFPC2) Planetary Camera (PC) F814W image originally presented in \cite{vanderMarel+1998AJ116.2220}. We fit the entire PC image. We adopt the point spread function appropriate for WFPC2/PC F814W, given in Table 3 of \cite{Cappellari+2002ApJ578.787}. To minimise the impact of extinction from the dust disc on our MGE model of the F814W image, we mask the north-western side of the dust disc, that appears to be in the foreground. We nevertheless include the central $9\times9\,$pixels to robustly constrain the stellar light in the galactic centre.

\begin{figure}
	\includegraphics[trim={2.5cm 0.5cm 2.5cm 1cm},width=0.95\columnwidth]{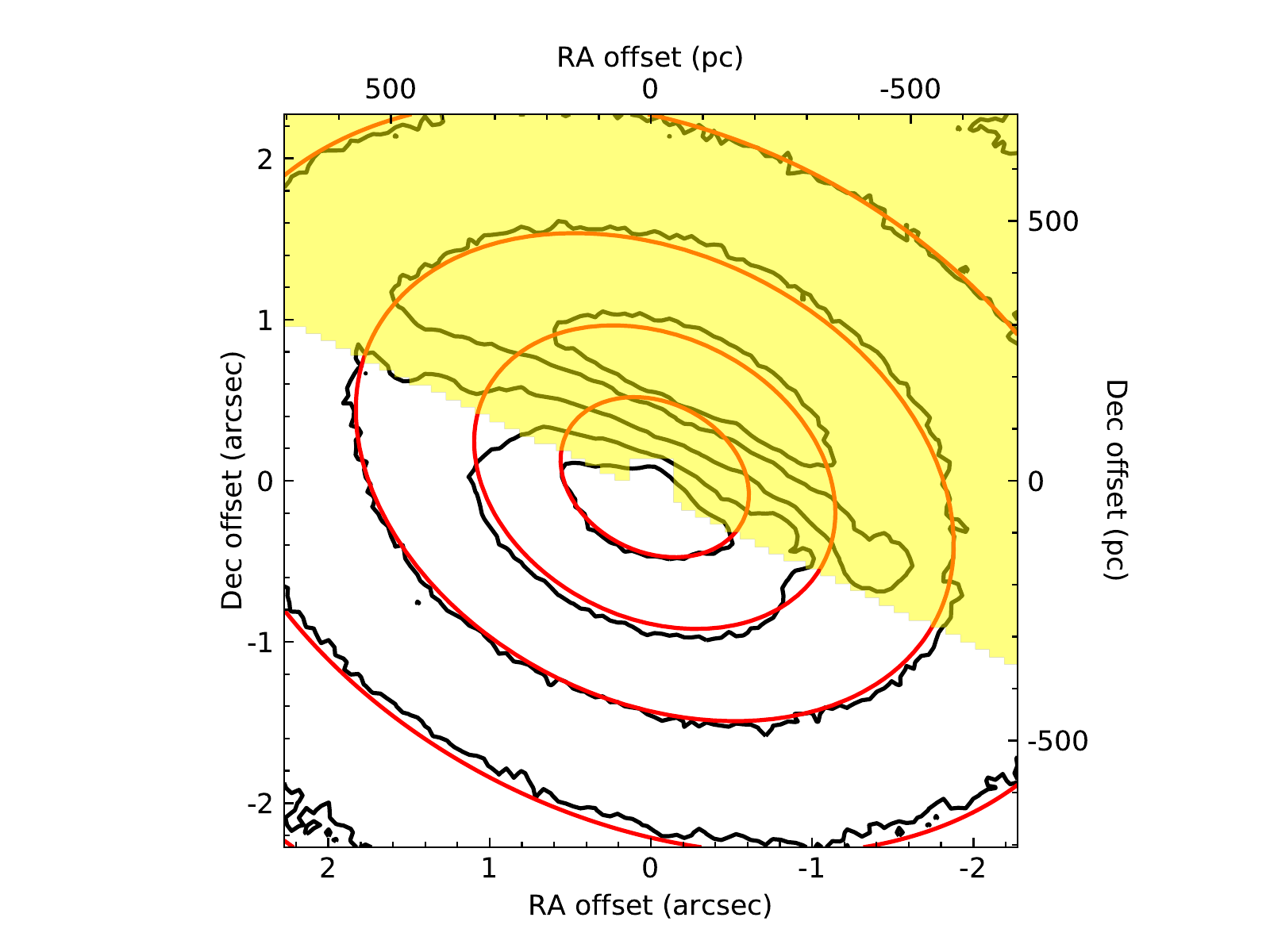}
    \caption{\textit{HST} WFPC2/PC F814W image of \mbox{NGC 7052} (black contours), overlaid with our MGE model (red contours). The north-western side of the image is masked (yellow) to exclude the foreground dust disc, but we retain the central $9\times9$ pixels.}
    \label{fig_mge}
\end{figure}

The MGE model consists of the deconvolved central intensity ($I^\prime$), width ($\sigma$) and apparent flattening ($q^\prime$) of a sequence of two-dimensional (2D) Gaussians that accurately replicate the observed (i.e. 2D, projected) light distribution. We convert these components to physical units ($I$-band solar luminosity surface densities $L_{\odot,I}\,$pc$^{-2}$) adopting a zero-point of $20.84\,$mag \citep{Holtzman+1995PASP107.1065} and an \textit{I}-band Solar absolute magnitude of $4.12\,$\citep{Willmer2018ApJS236.47}, both in the Vega system. These components are listed in Table \ref{tab_mge} and the fit is shown in Figure \ref{fig_mge}. The dust disc is evident in the distortions to the (otherwise elliptical) isophotes.

The MGE components describing the stellar light distribution can be converted into a mass distribution by multiplying the luminosity surface density of each Gaussian by the mass-to-light ratio. We assume this mass-to-light ratio to be radially constant, though we discuss this assumption further in Section \ref{ssec_bestModel}. Assuming an inclination, the projected stellar light (or mass) distribution can be analytically deprojected into a three-dimensional (3D) distribution, and the circular velocity resulting from this distribution can be calculated.

We will ultimately find that the stellar mass contribution to the potential within the central few resolution elements is negligible, and thus does not affect the best-fitting SMBH mass. This is corroborated by the spatially-resolved central Keplerian rotation curve, indicating that the central potential is dominated by a compact mass. In consequence, any extinction of the dust disc in the background of the south-eastern side of the galaxy does not significantly bias the inferred SMBH mass. 

\begin{table}
	\centering
	\caption{Spatially-deconvolved 2D MGE components of the model of our \textit{HST} WFPC2/PC F814W image of \mbox{NGC 7052}.}
	\label{tab_mge}
	\begin{tabular}{ccc} 
		\hline
		$\log_{10}{\left(\frac{I_j^\prime}{\mathrm{L}_{\odot,I}\,\mathrm{pc}^{-2}}\right)}$                              & $\log_{10}{\left(\frac{\sigma_j}{\mathrm{arcsec}}\right)}$               & $q_j^\prime$ \\
		(1)                                                             & (2)                                             & (3)                  \\
		\hline
		$4.49$                                                      & $-1.76\phantom{0}$                  & 0.73                \\
		$3.93$                                                      & $-0.23\phantom{0}$                  & 0.77		  \\
		$3.67$                                                      & $\phantom{-}0.14\phantom{0}$ & 0.69		  \\
		$3.56$                                                      & $\phantom{-}0.60\phantom{0}$ & 0.71		  \\
		\hline
	\end{tabular}
	\parbox{0.45\textwidth}{\textbf{Notes:} The table lists the central surface brightness (column 1), width (column 2) and axial ratio (column 3) of each deconvolved Gaussian component.}
\end{table}

\subsection{Molecular gas geometry}
In previous works in this series, we have commonly parametrized the molecular gas distribution using an axisymmetric exponential disc. The extremely high angular resolutions achieved with ALMA have however revealed that many objects host a central hole, that we have typically included using an additional central truncation. However, many galaxies have a molecular gas distribution which cannot be described by such a simple function. \cite{Smith+2019MNRAS485.4359} presented a new approach, using the \texttt{SkySampler} tool\footnote{\href{https://github.com/Mark-D-Smith/KinMS-skySampler}{https://github.com/Mark-D-Smith/KinMS-skySampler}} to infer the spatially deconvolved projected gas distribution (once reconvolved by the synthesised beam, this distribution is equivalent to the top-left panel of Figure \ref{fig_NGC7052_maps}), deproject this distribution into the disc plane under the thin disc assumption, and then calculate the associated line-of-sight velocities for the distribution as before. By construction, the model matches the observed gas distribution. The gas distribution therefore offers no constraint on the model parameters, but \texttt{SkySampler} allows us to remove a few degrees of freedom from the model. 

We adopt this approach for \mbox{NGC 7052}. The molecular gas surface brightness distribution appears to peak at a radius of ${\approx}0\farcs5$, before declining toward the centre of the galaxy (and outward). Attempting a fit using an exponential disc and central truncation failed to adequately reproduce the observed gas distribution. For our final fit, we therefore instead built a \texttt{SkySampler} model of the gas distribution from the projected \texttt{CLEAN} components, thus avoiding over-smoothing our model.

\subsection{Bayesian inference and priors}
The MCMC fit to our data explores the posterior probability distribution of our model, given by Bayes' theorem. Assuming uniform (maximum-ignorance) priors, and that our data has a Gaussian noise distribution constant for all pixels, the posterior is then proportional to the log-likelihood $(\ln P\propto-0.5\,\chi^2$), where the chi-squared goodness-of-fit statistic is given by
\begin{equation}
\label{eq_chisq}
\chi^2 \equiv \sum_i\left(\frac{\mathrm{data}_i - \mathrm{model}_i}{\sigma_i}\right)^2 = \frac{1}{\sigma^2}\sum_i(\mathrm{data}_i - \mathrm{model}_i)^2\,\,\,,
\end{equation}
where the sum is performed over all the pixels within the region of the data cube that the model fits, and $\sigma$ is the rms noise measured in line-free channels of the data cube.

Due to the very large number of constraints when fitting the entire 3D data cube, the ordinary assumption that the $1\sigma$ (67\%) confidence interval corresponds to $\Delta\chi^2\equiv\chi^2-\chi^2_\mathrm{min}=1$ (where $\chi^2_\mathrm{min}$ is the absolute $\chi^2$ minimum across all parameters explored) yields unrealistically small formal uncertainties. We therefore rescale the standard $\Delta\chi^2$ by a factor $\sqrt{2(N-P)}{\approx}\sqrt{2N}$, where $N{\approx}10^5$ is the number of constraints and $P=9$ is the number of free parameters of our model. This effectively rescales the uncertainties associated with our model parameters. This approach has been used in previous works of this series \citep[e.g.][]{Smith+2019MNRAS485.4359, North+2019MNRAS490.319} and other works encountering the same problem \citep[e.g.][]{vandenBosch+2009MNRAS398.1117, Mitzkus+2017MNRAS464.4789}. \cite{Smith+2019MNRAS485.4359} showed that this correction yields formal uncertainties that are consistent with those found by a bootstrap approach, and are thus more credible.

However, since adjacent pixels in our observations are not independent (i.e. the data are intrinsically spatially convolved by the synthesised beam, that is oversampled by our cube; see Table~\ref{tab_COdata}), failing to correct for pixel-to-pixel covariances would lead to underestimating the uncertainties. In previous works, we have corrected Equation \ref{eq_chisq} accordingly. The disadvantage of using this correction is that we need to introduce the inverse covariance matrix (with $N^2$ elements) to the calculated deviations, and in consequence can only fit a relatively small region of the cube. However, this correction is negligible compared to the $\sqrt{2N}$ rescaling described above, and so we neglect it in this work. This enables us to fit the entire molecular gas disc, rather than only some smaller central region as was previously necessary.

Finally, we impose physical bounds on each parameter to ensure the chain converges in a finite time, and that it does not explore unphysical regions of parameter space. Assuming maximal ignorance, we adopt uniform priors for all parameters except $M_\mathrm{BH}$ (see Table \ref{tab_MCMCresults}). As the SMBH mass can potentially span many orders of magnitude, we adopt instead a prior that is uniform in log-space for this single parameter, thus avoiding unduly favouring large values. 

\subsection{Best-fitting model}
\label{ssec_bestmodel}
\begin{figure*}
	\includegraphics[trim={2.5cm 0.5cm 2.5cm 1cm},width=0.95\textwidth]{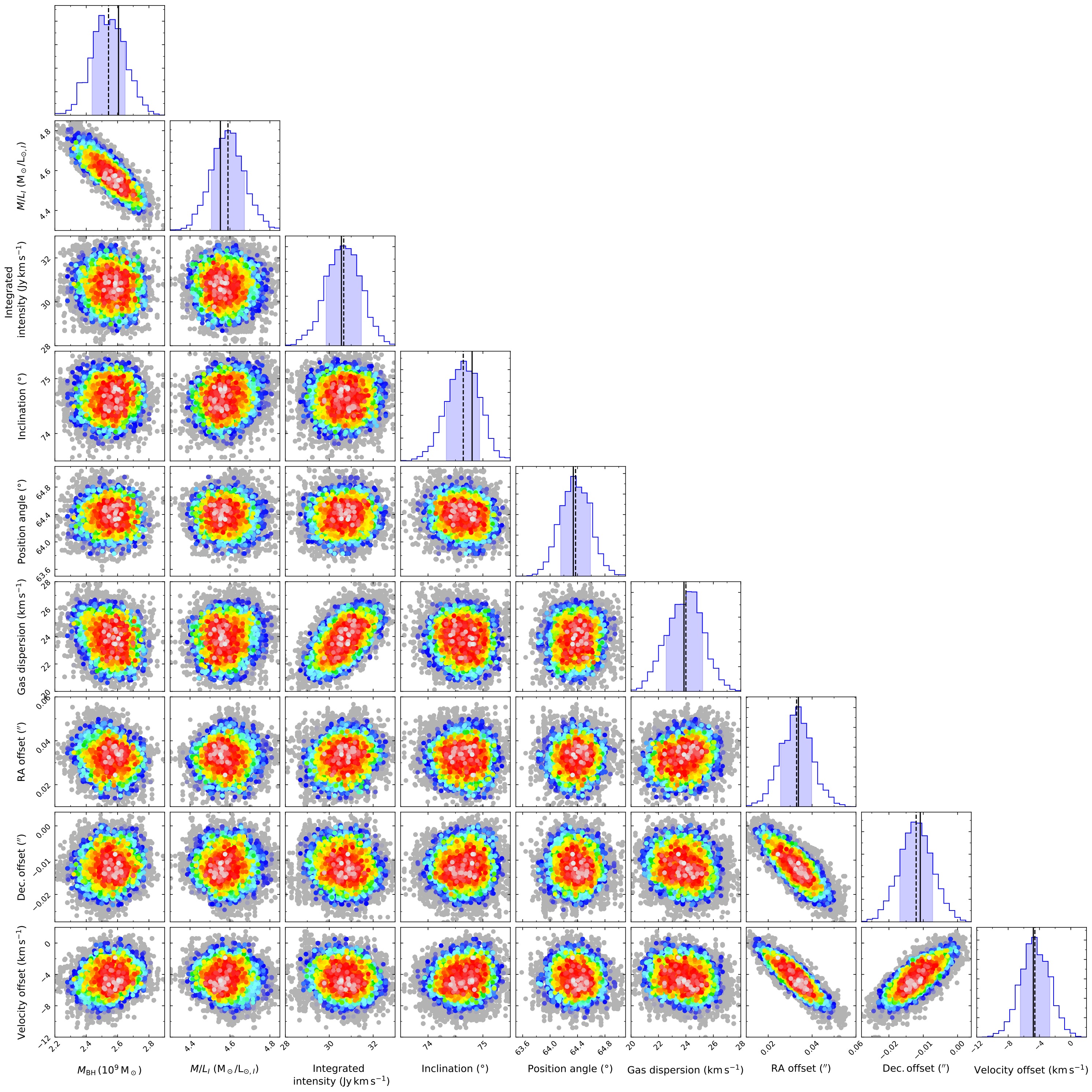}
    \caption{Corner plots showing the covariances between all model parameters, from our MCMC fit. Each point is a realisation of our model, colour-coded to show the relative log-likelihood of that realisation. Coloured points are within $\Delta\chi^2<\sqrt{2N}$ relative to the best-fitting model, with white points the most likely and blue points the least likely. Grey points are realisations with $\Delta\chi^2>\sqrt{2N}$, and are even less likely. The only significant physical covariance is between the SMBH mass and the mass-to-light ratio, that corresponds to attributing the same dynamical mass differently across the SMBH and stellar distribution. The covariances between the RA, Dec and velocity offsets correspond to moving the kinematic centre of the galaxy in three dimensions within a plane, and these offsets are much smaller than the resolution of our data. Each histogram shows the 1D marginalisation of a model parameter, the black lines denoting the median (dashed) and best-fitting (solid) values. The shaded region indicates the $68\%$ confidence interval. We note that the slight asymmetries of the posteriors imply that the most likely (best-fitting) and median value are very slightly different.}
    \label{fig_corner}
\end{figure*}

We ran our MCMC chain for $100~000$ steps, discarding the first $10~000$ steps as a burn-in. Our best-fitting model cube replicates the observed gas disc well. Figure \ref{fig_corner} shows the 2D marginalisation of each pair of input parameters, and the 1D marginalisation (histogram) of each parameter. As can be seen, all the 1D posteriors are approximately Gaussian, indicating the MCMC chain is well-converged. The coloured points in the 2D marginalisations indicate the log-likelihood of each model. The colour scale indicates points within $\Delta\chi^2<\sqrt{2N}$ of the best-fitting model, with white points the most likely (the best-fitting model is also shown by a solid black line in each histogram) and blue points the least likely. Grey points are realisations with $\Delta\chi^2>\sqrt{2N}$, and are even less likely. Slight asymmetries in the posterior, resulting from the highly non-linear model, imply that the median of each parameter is slightly different from the best-fitting parameter. However, both are consistent within the formal uncertainties for all parameters. The elliptical coloured `contours' also indicate that the posterior is well-sampled and well-converged.

The only significant physical covariance is the well-known one between the SMBH mass and the stellar mass-to-light ratio, equivalent to the conservation of total dynamical mass. The three offset parameters (centre right ascension, declination and velocity) are also correlated, as the gas disc is systematically distributed along a single plane in the cube. A small perturbation to one parameter will thus also change the other two to remain in this plane. 

The best-fitting, median, and formal uncertainties of each model parameter are listed in Table \ref{tab_MCMCresults}. The inferred SMBH mass is ($2.5\pm0.3)\times 10^9\,$\Msol$\,$ and $M/L_\mathrm{I}=(4.6\pm0.2)\,\mathrm{M_\odot}/\mathrm{L_{\odot,\textit{I}}}$, where both uncertainties are the $3\sigma$ (97\%) confidence level.

\begin{table*}
	\centering
	\caption{Best-fitting model parameters, with associated formal uncertainties.}
	\label{tab_MCMCresults}
	\begin{tabular}{lccccc} 
		\hline
		& \multicolumn{4}{c}{}\\
		Parameter & Priors & Best fit & Median & $1\sigma$ error & $3\sigma$ error\\
		(1) & (2) & (3) & (4) & (5) & (6) \\
		\hline
		\\
		Mass model: \\
		\hline
		SMBH mass ($10^9\,$\Msol) &  $10^5\;\rightarrow\;10^{12}$ & \phantom{0}2.61 &  \phantom{0}2.54 & 0.11 & 0.31\\
		Stellar $M/L_I$ ($\mathrm{M_\odot/L_{\odot,I}}$) & $\phantom{0}1\phantom{^{1}}\;\rightarrow\;10\phantom{^{11}}$ &  \phantom{0}4.55 &  \phantom{0}4.59 & 0.08 & 0.24\\
		\\
		Molecular gas disc:  \\
		\hline
                 $2\farcs5\times2\farcs5$ integrated intensity ($\mathrm{Jy\,}$\kms) & $\phantom{0}1\;\rightarrow\;200$ & 30.6\phantom{0} & 30.7\phantom{0} & 0.8\phantom{0} & 2.3\phantom{0} \\
		Gas velocity dispersion (\kms) & $\phantom{0}1\;\rightarrow\;100$ & 23.9\phantom{0} & 24.0\phantom{0} & 1.3\phantom{0} & 3.6\phantom{0}\\
		\\
		Viewing geometry: \\
		\hline		
		Inclination ($\degree$) & $60\;\rightarrow\;\phantom{0}89$ & 74.8\phantom{0} & 74.6\phantom{0} & 0.3\phantom{0} & 0.9\phantom{0}\\
		Position angle ($\degree$) & $\phantom{0}0\;\rightarrow\;359$ & 64.3\phantom{0} & 64.4\phantom{0} & 0.2\phantom{0} & 0.6\phantom{0}\\
				\\
		Nuisance Parameters: \\
		\hline
                 Centre RA offset ($\arcsec$) & $-0.1\;\rightarrow\;\phantom{0}0.1\phantom{01}$ & \phantom{-}0.034 & \phantom{-}0.033 & 0.007 & 0.021 \\
                 Centre Dec. offset ($\arcsec$) & $-0.1\;\rightarrow\;\phantom{0}0.1\phantom{01}$ & -0.011 & -0.012 & 0.005 & 0.014 \\
                 Centre velocity offset (\kms) &$\phantom{11\!}-75\;\rightarrow\;75\phantom{001111\!\!}$ & -4.8\phantom{00} & -4.6\phantom{00} & 1.9\phantom{00} & 5.4\phantom{00} \\
		\hline
	\end{tabular}
	\parbox[t]{0.95\textwidth}{\textbf{Notes:} Column 1 lists the input parameters of our dynamical model of NGC~7052. Column 2 lists the range allowed for each parameter, between which we adopt a uniform prior, except for the SMBH mass for which the prior is uniform in log-space. Column 3 lists the best-fitting parameter, while column 4 lists its median after marginalising over all other parameters. Columns 5 and 6 list the $1\sigma$ (67\%) and $3\sigma$ (99.7\%) confidence intervals of each parameter.}
\end{table*}

\section{Discussion}
\label{sec_discussion}
\subsection{Best-fitting mass model}
\label{ssec_bestModel}
\begin{figure*}
   \centering
   \includegraphics[trim={3.5cm 0.5cm 2.5cm 1cm},width=0.8\textwidth]{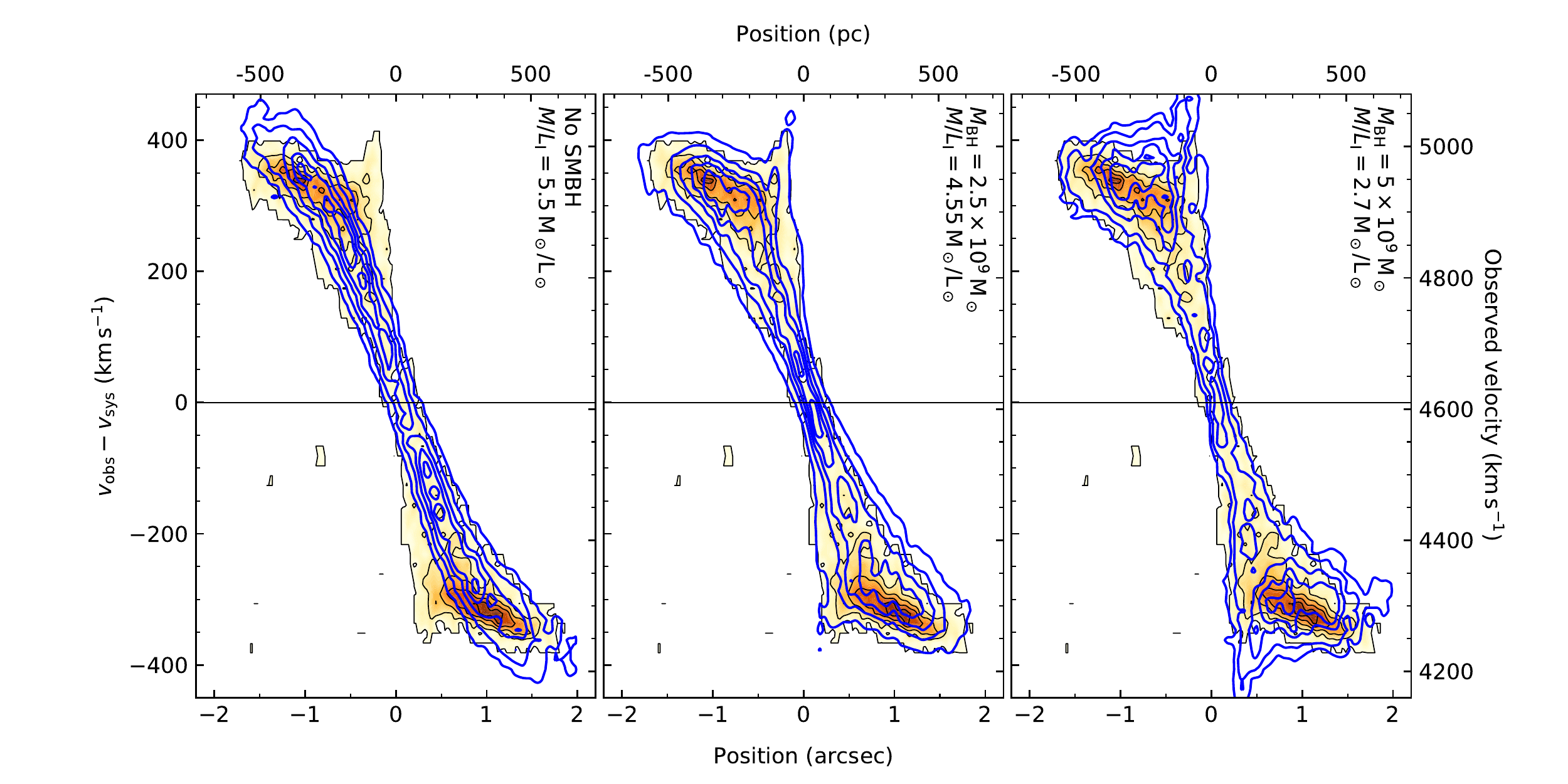}
   \caption{Model position-velocity diagrams along the kinematic major axis of \mbox{NGC 7052} (blue contours), showing a model without a SMBH (left), with the best-fitting SMBH (centre) and with an overly large SMBH (right). These are overlaid on the observed PVD (orange scales and black contours). The line-of-sight velocities at small radii are enhanced compared to those of a stellar mass-only model, thus requiring additional central mass to fully account for them.}
   \label{fig_BHPVD}
\end{figure*}

The quality of our best-fitting model is easy to assess from a kinematic major-axis PVD, as shown in Figure \ref{fig_BHPVD}, although it should be noted that our fit was performed to the entire data cube, not only to this PVD. The left panel shows a fit to the observed data cube assuming no SMBH. To attempt to account for the high velocities observed at small radii, the fit adopts a larger $M/L_\mathrm{I}$, however this is clearly not a good match to the observations. The right panel shows another fit assuming a SMBH mass larger than that found in our best model. The fit attempts to compensate by reducing $M/L_\mathrm{I}$, however again this yields a poor fit. The central panel clearly shows that our best model recovers the observed Keplerian rotation within the central region dominated by the SMBH, and the asymmetry of this signature on either side of the disc. Since the only non-axisymmetric feature of our model is the gas distribution, it is clear that the observed asymmetry is the result of the lack of gas to properly sample the Keplerian rise on the south-western (negative velocities) side of the disc, rather than evidence of disturbed motions. 

The velocity field residuals, obtained by subtracting the model velocity field from that shown in Figure \ref{fig_NGC7052_maps}, show no spatial structure that would indicate organised non-circular motions (as were found in e.g. \citealt{Smith+2019MNRAS485.4359}). In addition, the very low velocity dispersions indicate that the gas velocities are dominated by circular motion. Throughout the disc, $v/\sigma{\approx}15$ (where $v$ is the deprojected rotation velocity and $\sigma$ the intrinsic velocity dispersion), indicating that the gas is rotationally-supported. 

In principle, the stellar mass-to-light ratio can vary across the galaxy, tracing changes of the stellar population \citep[e.g.][]{Davis+2017MNRAS464.453, Davis+2018MNRAS473.3818}. No such variation is required to adequately fit our data, but as always a sudden change in the mass-to-light ratio in the centre of the galaxy could obviate the need for a SMBH. There is no photometric evidence to support such a change, and the variation required would be unphysically large - a factor of ${\approx}50$. 

\subsection{Systematic uncertainties}
\label{ssec_systematics}
SMBH mass uncertainties due to the inclination scale as \mbox{$M_\mathrm{BH} \propto 1/\sin^2 i$} \cite[e.g.][]{Smith+2019MNRAS485.4359}. At low inclinations, the inclination uncertainty can dominate the SMBH mass uncertainty. At the highest inclinations, other effects become important, such as the inability to resolve non-axisymmetric structures, the disc's intrinsic thickness along any line of sight, and potentially the gas optical depth, all of which lessen the accuracy of a dynamical model. The molecular gas disc in \mbox{NGC 7052} is reasonably highly inclined ($i{\approx}70\degree$) and has very small inclination uncertainties which make only a very small contribution to the total $M_\mathrm{BH}$ uncertainty budget. Indeed, simulations suggest that $i{\approx}70\degree$ appears to be an optimal inclination for accurately recovering SMBH masses from molecular gas kinematics (\citealt{Davis2014MNRAS443.911}).

Inaccuracies in the mass model adopted can, in general, also bias the recovered SMBH mass, as an incorrect share of the dynamical mass is assigned to the SMBH. Beside the SMBH, our mass model includes only a contribution from the stellar mass distribution, and it neglects both gas and (dark) halo contributions. However, the relevant length scale on which these contributions matter is that traced by the CO disc, that extends only to a radius of ${\approx}1\farcs5$. Over such a small scale, dark matter likely makes a negligible contribution to the overall mass budget. Contributions from warm gas ($10^{3.6}\,$\Msol$\,$in total; \citealt{Pandya+2017ApJ837.40}) and the dust disc ($10^{4}\,$\Msol$\,$in total; \citealt{Nieto+1990AandA235.L17}) are similarly negligible. Naturally, if any of these components were radially distributed identically to the stellar mass, their only effect would in any case be to change the derived dynamical mass-to-light ratio. A radially-varying distribution would lead to a mass-to-light ratio gradient, but it would require a significantly centrally-concentrated mass distribution to substantially affect the derived $M_\mathrm{BH}$.

\begin{figure}
   \centering
   \includegraphics[trim={1.5cm 0.5cm 2.5cm 1cm},width=0.9\columnwidth]{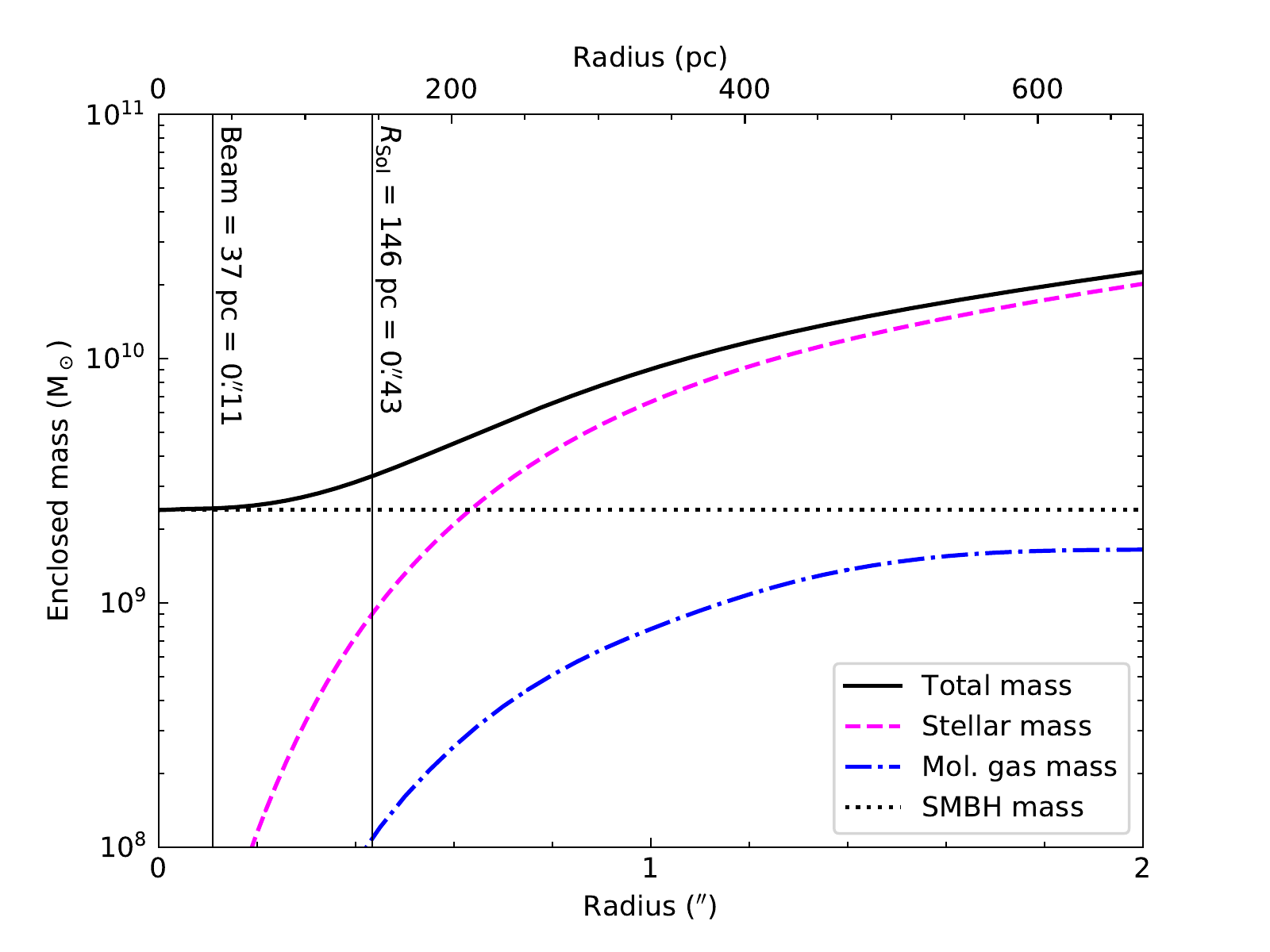}
   \caption{Cumulative mass function of \mbox{NGC 7052}, showing the relative contributions from the SMBH (black dotted line), molecular gas (blue dot-dashed line) and stars (violet dashed line). The total enclosed mass is shown by the solid black line. The physical scales of the synthesised beam and SMBH sphere of influence (assuming our measured SMBH mass and a stellar velocity dispersion $\sigma_\mathrm{e}=266\,$\kms) are indicated by vertical black lines. The molecular gas mass contribution is negligible at all radii.}
   \label{fig_massFunc}
\end{figure}

Figure \ref{fig_massFunc} shows the enclosed mass of our best-fitting model within spheres of increasing galactic radii, with the contributions from the SMBH, stars and molecular gas indicated separately. Also indicated are the radii corresponding to the synthesised beam and $R_\mathrm{SoI}$, the latter using our best-fitting $M_\mathrm{BH}$ and $\sigma_\mathrm{e}=266\,$\kms \citep{Gultekin+2009ApJ698.198}. As is clearly seen, the SMBH dominates the galactic potential not only within its nominal sphere of influence, but up to ${\approx}0\farcs6$ (${\approx}200\,$pc). We thus resolve this region radially with approximately 6 beams. The molecular gas contribution is negligible at all radii. 

We note that the radius at which the SMBH and stars have equal contributions ($R_\mathrm{eq}$) is around $60\%$ larger than the nominal SMBH sphere of influence. This is not necessarily concerning, as typical early-type galaxies have $R_\mathrm{eq}$ slightly larger than $R_\mathrm{SoI}$ \citep{Yoon2017MNRAS466.1987}.

Next, we consider the accuracy of our adopted stellar mass model. Although the MGE model appears to match well the \textit{HST} F814W image over the centre of the galaxy (Figure \ref{fig_mge}), this region is strongly affected by dust. Dust attenuation is expected to decrease the observed flux, and hence cause us to attribute too little mass to the stellar contribution, potentially overestimating $M_\mathrm{BH}$. We argue that this effect can be safely disregarded here, as it has been carefully mitigated. Firstly, we adopted the \textit{HST} F814W image of the galaxy to build our stellar light model. We also masked the north-western side of the dust disc, where it is in the foreground. Adopting this relatively long wavelength, and masking the foreground dust, will reduce the extinction. Secondly, as we have argued previously, an erroneous stellar light profile can be corrected by an appropriate change to the mass-to-light ratio. Thus, inferring the mass-to-light ratio from beyond the dust disc and assuming it is radially constant would significantly bias our results if no correction was made for dust extinction. However, our stellar mass-to-light ratio is determined purely by the CO kinematics, that only extends across the dust disc. Assuming the extinction due to this disc does not vary dramatically, the effect on the stellar light model will be compensated by an associated change in the mass-to-light ratio. In Section \ref{ssec_bestModel}, we have further shown that there is no evidence for a mass-to-light ratio gradient, that would be a consequence of a substantial deviation of the photometrically-derived stellar light profile and the dynamically-derived mass profile.

In any case, due to the very high spatial resolution of our data, we probe well into the SMBH-dominated regime, where the stellar contribution is small (see Figure \ref{fig_massFunc}). We therefore conclude that any remaining uncertainties in our stellar light model will not significantly bias our SMBH mass.

We assumed in Section \ref{sec_dynamics} that the CO disc is razor-thin, implemented by setting the $z$-coordinate (that orthogonal to the disc plane) of the \texttt{KinMS} particles to zero. Notionally, a non-negligible disc thickness could account for some of the observed line width along each line of sight, reducing the intrinsic gas velocity dispersion required. To test this assumption, we run another MCMC chain instead giving each particle a $z$-position drawn from a uniformly-distributed radially-constant disc thickness of $\pm d$, where $d$ is an additional free model parameter. We adopt a uniform prior of $0<d<3.3\,$kpc (this upper bound far larger than the disc scale). This chain yields a disc thickness consistent with the synthesised beam, with negligible improvement in the associated best-fitting model's log-likelihood. The associated best-fitting SMBH mass and stellar mass-to-light ratio are unchanged. In addition, the best-fitting gas velocity dispersion found by the new model is not smaller than that found assuming a thin disc by a statistically-significant factor. We therefore conclude that the thin disc assumption is acceptable when interpreting data at our resolution, though higher-resolution observations may prove otherwise.

Finally, the adopted distance to \mbox{NGC 7052} sets the scale of our dynamical model. The inferred SMBH mass scales linearly with distance, since $M_\mathrm{BH} \propto v^2R \propto D$, where $v$ is the rotation velocity of a particle at radius $R$ (as we observe an angular radius, the physical radius scales with the assumed distance).

We have adopted a distance of $69.3\,$Mpc for consistency with the MASSIVE survey \citep{Ma+2014ApJ795.158}. Although \cite{Ma+2014ApJ795.158} do not quantify the uncertainty of this distance, the Hubble flow distances listed in the NASA/IPAC Extragalactic Database\footnote{\href{http://ned.ipac.caltech.edu/}{http://ned.ipac.caltech.edu}} have a typical uncertainty of $7\%$. As is standard practice, we do not include this uncertainty in our quoted dynamical SMBH mass measurement, and the results herein can simply be corrected to any adopted distance. 

\subsection{Gas velocity dispersion}
\label{ssec_vdisp}
The line-of-sight velocity dispersions observed in molecular gas are comprised of an intrinsic (turbulent) velocity dispersion, broadened by beam smearing of mean velocity gradients. Typical molecular gas intrinsic velocity dispersions are very small (often ${<}10\,$\kms; e.g. \citealt{Davis+2017MNRAS468.4675, Davis+2018MNRAS473.3818, Smith+2019MNRAS485.4359}). 

\cite{vanderMarel+1998AJ116.2220} found that the H$\alpha$ velocity dispersion of \mbox{NGC 7052} decreased with increasing radius, with a central peak of $400\,$\kms$\,$ falling to $70\,$\kms$\,$ by a radius of ${\approx}1\arcsec$. Although enhanced central dispersions are expected by Doppler broadening close to the central SMBH, a model excluding an intrinsic velocity dispersion gradient was inconsistent with their observations \citep{vdBosch+1995MNRAS274.884}. In their dynamical models, they found that an exponentially-decaying intrinsic (turbulent) velocity dispersion was required to account for the above variation, of the form 
\begin{equation}
\sigma(R) = \sigma_0 + \sigma_1 e^{-R/R_\mathrm{t}}\,,
\label{eq_vdisp_vdMarel}
\end{equation}
where $R_\mathrm{t}$ is the scale length of the (turbulent) velocity dispersion and $\sigma_0$ and $\sigma_1$ parametrize the radial variation. Their best-fitting dynamical model yielded $\sigma_0=60\,$\kms, $\sigma_1=523\,$\kms$\,$ and $R_\mathrm{t}=0\farcs11$. The very small scale length implies that although the central amplitude is large, the dispersion is dominated at almost all radii by the (rather large) constant term.

Our best-fitting model described in Section \ref{sec_dynamics} assumed a radially-constant velocity dispersion. For comparison, we performed another fit allowing the velocity dispersion to vary with radius according to Equation \ref{eq_vdisp_vdMarel}. This model is visibly inferior to that found assuming a constant dispersion, but the best-fitting SMBH mass is consistent with our previous result. We therefore conclude that no intrinsic velocity dispersion gradient is required to account for our observations, and our derived SMBH mass is robust.

\subsection{Comparison with ionised gas}
\cite{vanderMarel+1998AJ116.2220} used \ion{H}{$\alpha$} and [\ion{N}{II}] emission observed with the \textit{HST} Faint Object Spectrograph to measure the central SMBH mass of \mbox{NGC 7052}, and found \mbox{$M_\mathrm{BH}=3.9^{+2.7}_{-1.5}\times10^8\,$\Msol} (corrected to our adopted distance). Our measurement is not consistent with this result. 

\begin{figure}
   \centering
   \includegraphics[trim={2.5cm 0.5cm 2.5cm 1cm},width=0.8\columnwidth]{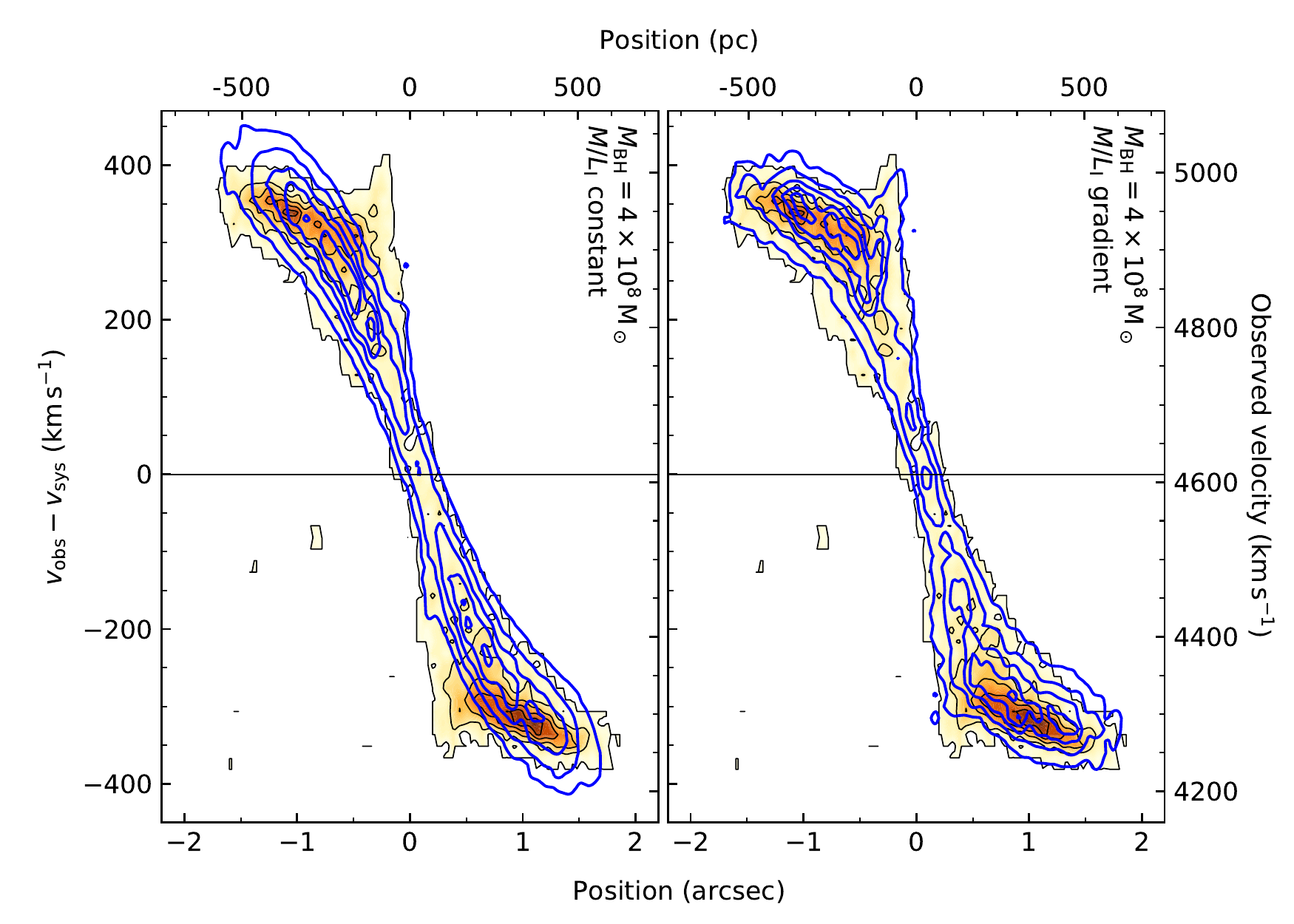}
   \caption{Model position-velocity diagrams along the kinematic major axis of the galaxy (blue contours), showing the best-fitting models with the SMBH mass fixed to that of \protect\cite{vanderMarel+1998AJ116.2220}, with either a radially-constant mass-to-light ratio (left), or a mass-to-light ratio gradient (right). These are overlaid on the observed PVD (orange scales and contours). Although allowing a mass-to-light ratio gradient helps to enhance gas velocities at small radii, this model remains inferior to that described in Section \ref{ssec_bestmodel}.}
   \label{fig_gasBHPVD}
\end{figure}

As a check, we performed another fit to our observations, with the SMBH mass set to that found by \cite{vanderMarel+1998AJ116.2220} from warm gas kinematics. The major-axis PVD of the model with the maximum log-likelihood is shown in Figure \ref{fig_gasBHPVD} (left panel), overlaid on our ALMA data. Clearly, the model severely underestimates the molecular gas velocities at small radii, as would be expected from imposing a SMBH mass one-quarter of that required. 

This can be partially compensated for by allowing a stellar mass-to-light ratio gradient. We thus include a gradient in yet another model by calculating the circular velocity as before, but assuming $M/L_I=1\,\mathrm{M_\odot/L_{\odot,\textit{I}}}$, and then multiplying this function at each radius by $\sqrt{M/L_I(R)}$ \citep{Davis+2017MNRAS464.453, Davis+2018MNRAS473.3818}. We adopt a linearly-varying mass-to-light ratio that flattens beyond $2\arcsec$. The best-fitting model is shown in Figure \ref{fig_gasBHPVD} (right panel), and has a central mass-to-light ratio of $6.9\,\mathrm{M_\odot/L_{\odot,\textit{I}}}$, returning to $4.6\,\mathrm{M_\odot/L_{\odot,\textit{I}}}$ (our best-fitting spatially-constant $M/L_\textit{I}$) at $2\arcsec$.  Although as expected the M/L gradient increases the central velocities, the model is still inferior to that presented in Section \ref{ssec_bestModel}. As we discussed in Section \ref{ssec_systematics}, a discrete increase in the mass-to-light ratio at very small (spatially-unresolved) scales can always mimic a SMBH signature, but there is no physical reason to expect such a change. We therefore conclude here that such a mass-to-light ratio gradient is disfavoured, and hence that the $M_\mathrm{BH}$ measurement of \cite{vanderMarel+1998AJ116.2220} is excluded by our data.

The main advantages of our molecular gas observations over those used by \cite{vanderMarel+1998AJ116.2220} are as follows. First, our observations trace the entire gas disc, rather than only a few discrete locations along the major axis (the galactic radii of which can themselves have significant uncertainty due to pointing uncertainty). By fitting the entire gas disc, we have many more constraints on the observed kinematics (and the uncertainty on their locations), and hence on the mass distribution throughout the central region of the galaxy. Second, all gas dynamical measurements can be affected by non-gravitational forces and non-circular motions. The very low velocity dispersions of our CO gas indicate that these are negligible (while warm ionised gas is likely to be more significantly affected). As outlined in Section \ref{ssec_vdisp}, \cite{vanderMarel+1998AJ116.2220} required a significant central velocity dispersion to adequately fit their observations, attributed to turbulence and neglected in the dynamical model. If this dispersion instead corresponds to (some component of) pressure support, the fit will necessarily underestimate the SMBH mass. It should be further noted that more recent \textit{HST} Space Telescope Imaging Spectrograph observations indicate the presence of a separate ionised-gas dynamical component (perhaps a broad-line region; \citealt{NoelStorr+2003ApJS148.419,NoelStorr+2007ApJ663.71,VerdoesKleijn+2006AJ131.1961}), that may exhibit significantly different kinematics to that of the extended gas detected by the FOS.

A similar case of a rotating warm gas disc with a strong velocity dispersion gradient is found in Centaurus A. \cite{HaringNeumayer+2006ApJ643.226} explored the sensitivity of SMBH masses inferred from dynamical models to the inclusion of the velocity dispersion gradient as component of the dynamical support. They found that a cold disc assumption could underestimate the SMBH mass by a factor of ${\approx}3$ in their case, with respect to a model including the velocity dispersion gradient. Although the degree to which the lack of this support can underestimate the SMBH mass will vary between discs, this evidence suggests that the lack of dynamical pressure support in the warm gas model of \cite{vanderMarel+1998AJ116.2220} could be the reason for the disagreement between their inferred SMBH mass and ours.

The $M_\mathrm{BH}{-}\sigma_\mathrm{e}$ relation of \cite{Sahu+2019ApJ887.10} predicts \mbox{$M_\mathrm{BH}=1.0^{+2.1}_{-0.7}\times10^9\,$\Msol$\,$} for NGC~7052 (assuming $\sigma_\mathrm{e}=266\pm13\,$\kms, and including $0.44\,$dex of intrinsic scatter). Our result is in excellent agreement with this prediction, whereas the ionised-gas measurement of \cite{vanderMarel+1998AJ116.2220} is significantly below it. The significant differences across SMBH masses derived via different dynamical tracers thus continues to demonstrate the need for robust cross-checks between all techniques. Further SMBH mass measurements using molecular gas offer the prospect of determining the intrinsic scatters of the SMBH-host galaxy scaling relations with measurements from a single technique across the entire Hubble sequence.

\subsection{Tidal accelerations and molecular cloud stability in the galactic centre}
\label{ssec_shear}
The molecular gas discs of many galaxies in the WISDOM sample exhibit central holes at small radii \citep[e.g.][]{Davis+2018MNRAS473.3818, Smith+2019MNRAS485.4359}, including \mbox{NGC 7052}. These ${\approx}100\,$pc holes have been revealed for the first time by the exceptionally high angular resolutions required for SMBH measurements. The typical extents of these features are roughly consistent with the SMBH spheres-of-influence, suggesting that they may have a dynamical origin. 

One dynamical mechanism that could give rise to depleted molecular gas surface densities at the centre of galaxies is the tidal disruption of gas clouds. It is generally believed that molecular gas forms in these clouds, due to the outer layers of the clouds shielding their centres from ultraviolet radiation that would otherwise photo-dissociate the molecules, and due to the high densities increasing the number of collisions that can form molecules (and those with dust grains that can enhance molecule formation through surface reactions; \citealt{BinneyMerrifield1998book}). 
Strong shear or tidal acceleration could exceed the self-gravity of such clouds, disrupting them and exposing the molecules to photo-dissociation, or preventing the formation of clouds entirely. This would in turn inhibit the formation of stars near an SMBH \citep[e.g.][]{Sarzi+2005ApJ628.169}.

\begin{figure}
   \centering
   \includegraphics[trim={3cm 1.5cm 3cm 1cm},width=0.55\columnwidth]{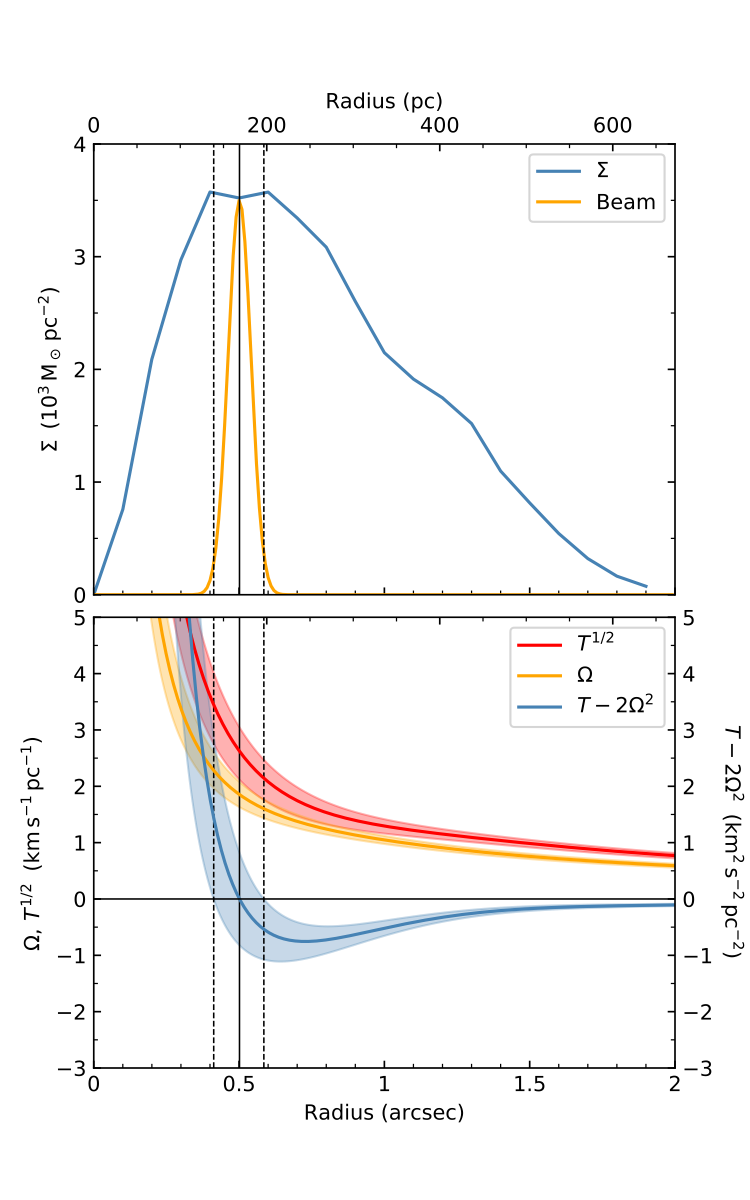}
   \caption{\textbf{Top panel:} Azimuthally-averaged molecular gas surface density radial profile ($\Sigma$; blue solid line), overlaid with our synthesised beam (orange solid line) centred at $0\farcs5$. \textbf{Bottom panel:} Orbital angular velocity ($\Omega$, orange solid line), tidal acceleration per unit length in the radial direction ($T$, red solid line) and the function $T-2\Omega^2$ (blue solid line), all calculated from our best-fitting dynamical model. Coloured envelopes around each line indicate the $\pm3\sigma$ confidence intervals.  $T-2\Omega^2$ is positive within $0\farcs50\pm0\farcs09$, indicated by a black vertical line (with $3\sigma$ confidence intervals indicated by black dashed vertical lines) in both panels. This matches well the maximum of the surface density profile, and thus the radius within which the molecular gas density rapidly decreases.}
   \label{fig_shearPlot}
\end{figure}

\cite{Liu+2020MNRASinrev} considered the effect of external gravity on the morphology and confinement of giant molecular clouds. In their formalism, spatial variations of the external gravitational potential can contribute to either keeping clouds bound or to disrupting them, depending on the sign of $T-2\Omega^2$, where 
\begin{equation}
T(R) \equiv -R \frac{d\Omega^2(r)}{dr}\bigg|_R
\end{equation}
is the tidal acceleration in the radial direction and $\Omega$ is the orbital angular velocity ($v/R$; see Appendix A of \citealt{Liu+2020MNRASinrev}).
These quantities, derived from our best-fitting dynamical model, are shown in the bottom panel of Figure \ref{fig_shearPlot}. Uncertainties in each are estimated by propagating the uncertainties in our model parameters via Monte Carlo methods.

Our model indicates that $T-2\Omega^2$ changes sign at $0\farcs50\pm0\farcs09$ and is positive (thus disrupting the clouds) within this radius. This position is consistent with the peak of the gas distribution (Figure \ref{fig_shearPlot}, top panel). If other contributions to the energy budgets of clouds at these radii are negligible (or, more likely, are finely balanced by gravity), the central gas deficit could be the result of tidal accelerations disrupting the clouds. We cannot directly measure these other contributions in \mbox{NGC 7052}, and thus cannot robustly test this hypothesis.

Entirely different explanations are of course also possible. Emission from a central AGN could contribute sufficient photons to dissociate the CO molecules. Alternatively, holes may be better traced by higher-$J$ CO transitions \citep[e.g.][]{GarciaBurillo+2016ApJL823.12}, or by dense molecular gas emission \citep[e.g.][]{Imanishi+2018ApJL853.25}. The holes found thus far by the WISDOM project (\citealt{Davis+2018MNRAS473.3818, Smith+2019MNRAS485.4359} and this work, plus a slight central depression in \citealt{North+2019MNRAS490.319}) do not appear to be correlated with AGN activity. Another dynamical possibility is that resonances due to non-axisymmetric features in the potential could cause the central hole. \cite{Davis+2018MNRAS473.3818} investigated this for NGC~4429 and concluded that an unusually fast pattern speed would be required, making this explanation unlikely.

\section{Conclusions}
\label{sec_conclusions}
High angular resolution observations from the Atacama Large Millimetre/sub-millimetre Array (ALMA) and Atacama Compact Array (ACA) were used to make a $1.3\,$mm continuum image and a $^{12}$CO(2-1) cube of the elliptical galaxy \mbox{NGC 7052}. We detect a compact continuum source at the optical centre of the galaxy, assumed to correspond to emission from the active galactic nucleus. The CO data reveal a dynamically cold ($\sigma{\approx}20\,$\kms) rotating disc coincident with a prominent dust disc visible in \textit{Hubble Space Telescope} (\textit{HST}) images. The ALMA observations resolve a physical scale of $0\farcs11$ ($37\,$pc), smaller than the central region over which the galactic gravitational potential is dominated by the central supermassive black hole (SMBH).

We constructed a dynamical model of \mbox{NGC 7052} to constrain the SMBH mass. We estimated the stellar contribution to the potential by multiplying a multi-Gaussian expansion of a \textit{HST} WFPC2/PC F814W optical image by a spatially-constant mass-to-light ratio. The model was fit to the central $2\farcs56\times2\farcs56$ region of the ALMA data cube within a Markov Chain Monte Carlo framework. The inferred SMBH mass is $(2.5\pm0.3)\times10^9\,\mathrm{M_\odot}$ and the \textit{I}-band mass-to-light ratio is $(4.6\pm0.2)\,\mathrm{M_\odot/L_{\odot,I}}\,$ ($3\sigma$ confidence intervals). We exclude the possibility of a physically-motivated mass-to-light ratio gradient.

This SMBH mass measurement is substantially larger than that found previously using \textit{HST} Faint Object Spectrograph observations of ionised gas by \cite{vanderMarel+1998AJ116.2220}. The key difference is that the molecular gas disc is dynamically cold even very close to the SMBH, whereas the warm gas kinematics of \cite{vanderMarel+1998AJ116.2220} show large velocity dispersion gradients. Our observations strongly exclude their previous measurement. We suggest that our larger SMBH mass measurement is due to the fact that they did not include dynamical pressure support in their models. 

The peak molecular gas surface density occurs at a radius of ${\approx}0\farcs5$, the surface density slowly declining towards the centre of the galaxy (and outward). This peak corresponds to the radius within which the external gravitational potential acts to tidally disrupt molecular gas clouds. We suggest that if this effect dominates the self-gravity of clouds, it is likely that the central molecular gas depletion is the result of tidal forces preventing the formation of molecular clouds.

Our SMBH measurement once more demonstrates the power of the molecular gas kinematics method to accurately measure SMBH masses, and the important role ALMA can play to understand the dynamics of molecular gas in the central regions of galaxies. The steadily increasing sample of such masses will soon allow us to constrain the $M_\mathrm{BH}{-}\sigma_\mathrm{e}$ relation over several orders of magnitude in SMBH mass with a single method.

\section*{Acknowledgements}
MDS acknowledges support from a Science and Technology Facilities Council (STFC) DPhil studentship under grant ST/N504233/1. MB was supported by STFC consolidated grant `Astrophysics at Oxford' ST/H002456/1 and ST/K00106X/1. TAD was supported by STFC consolidated grant ST/S00033X/1. MC acknowledges support from a Royal Society University Research Fellowship. TGW acknowledges funding from the European Research Council (ERC) under the European Union’s Horizon 2020 research and innovation programme (grant agreement No. 694343).

This paper makes use of the following ALMA data: 
ADS/JAO.ALMA
\#2016.2.00046.S and
\#2018.1.00397.S.
ALMA is a partnership of ESO (representing its member states), NSF (USA) and NINS (Japan), together with NRC (Canada), MOST and ASIAA (Taiwan), and KASI (Republic of Korea), in cooperation with the Republic of Chile. The Joint ALMA Observatory is operated by ESO, AUI/NRAO and NAOJ.

This research has made use of the NASA/IPAC Extragalactic Database (NED), which is operated by the Jet Propulsion Laboratory, California Institute of Technology, under contract with the National Aeronautics and Space Administration. This paper has also made use of the HyperLeda database (\href{http://leda.univ-lyon1.fr}{http://leda.univ-lyon1.fr}).

\section*{Data Availability}
The observations underlying this article are available in the ALMA archive, at \href{https://almascience.eso.org/asax/?result\_view=observation\&sourceNameResolver=NGC7052\&projectCode=2016.2.00046.S\%7C2018.1.00397.S}{https://almascience.eso.org/asax/}, and in the Hubble Legacy Archive, at \href{https://hla.stsci.edu/}{https://hla.stsci.edu}.




\bibliographystyle{mnras}
\bibliography{papers} 




%
%


\bsp	
\label{lastpage}
\end{document}